\documentstyle[12pt]{article}
\def\anp#1#2#3{
        Ann. Phys. (N.Y.) {\bf #1 }, #2 (19#3)}

\def\phl#1#2#3{
        Phys. Lett. {\bf #1}, #2 (19#3)}
\def\prl#1#2#3{
        Phys. Rev. Lett. {\bf #1}, #2 (19#3)}

\def\phr#1#2#3{
        Phys. Rep. {\bf #1}, #2 (19#3)}
\def\prd#1#2#3{
        Phys. Rev. D {\bf #1}, #2 (19#3)}
\def\nup#1#2#3{
        Nucl. Phys. {\bf #1}, #2 (19#3)}

\def\beq{\begin{equation}}
\def\eeq{\end{equation}}
\def\zn{$Z(N)\;$}
\def\sun{$SU(N)\;$}
\begin{document}
\begin{flushright}
BNL--P--2/92\\
December, 1992\\
\end{flushright}
\vspace{.5in}
\begin{center}
{\large \bf Medley in finite temperature field theory} \\
\vspace{.5in}
Robert D. Pisarski\\
{\it Department of Physics,Brookhaven National Laboratory\\
Upton, New York  11973   USA}\\
\end{center}
\vspace{.125in}
\begin{center}
Abstract
\end{center}
\begin{quotation}
I discuss three subjects in thermal field theory:
why in \sun gauge theories the \zn symmetry is broken at high
(instead of low) temperature,
the possible singularity structure of gauge variant
propagators,
and the problem of how to compute the viscosity from the Kubo formula.
\end{quotation}
\begin{center}
{\it Based on an invited talk presented at the
Workshop on Hot QCD held at the Winnipeg Institute for
Theoretical Physics, Winnipeg, July 1992.}
\end{center}

\vspace{.25in}

I take this opportunity to make three comments about
field theories at nonzero temperature,
especially as applies to hot $QCD$.  The first is trivial;
the second, base speculation, and the third, a technical problem.

\vspace{.25in}
\noindent{\bf 1. Why the \zn symmetry is broken at high temperature}
\vspace{.125in}

Consider an \sun
gauge theory without dynamical fermions.  At nonzero temperature
the order parameter
for this system is the Polyakov line [\ref{r1}],
$$
L(x) =
 tr \left( {\cal P} \, exp\left( ig \int^\beta_0 A_0(x) d\tau
\right) \right) \, ;
$$
${\cal P}$ refers to path ordering,
$g$ is the coupling constant, and $\beta = 1/T$, with $T$ the
temperature.  The Polyakov line is invariant under
gauge transformations which are strictly periodic
in the euclidean time $\beta$.
There are also gauge transformations
which are periodic only up to an element of the center of the group [\ref{r2}];
the aperiodicity is a global \zn phase,
$=exp( i \phi_j)$, $\phi_j = 2 \pi j/N$ with
$j = 0...(N-1)$.  The Polyakov line transforms under these
aperiodic gauge transformations as
$L(x) \rightarrow exp(i \phi_j) L(x).$
The Polyakov line has a nontrivial vacuum
expectation value at high temperature,
choosing one of $N$ equivalent vacua, and vanishes below the transition
temperature.
This behavior is exactly opposite to that of a typical spin system,
where the symmetry is broken at low temperature, and restored at high
temperature.  Here I give a simple argument why.

In a spin system with discrete symmetry, the simplest argument for
spontaneous symmetry breaking at low temperature is due to Peierls [\ref{r3}].
Consider an Ising model, with a spin $= \pm$ on each site, and a coupling
$J$ between adjacent sites, so the hamiltonian $= - J \sum_{i,j} \sigma_i
\cdot \sigma_j$.  Start with a perfectly ordered state, where
all spins = $+$.  Now insert a domain of width
$\sim R$ into this state, where all spins
$= -$ inside the domain.  Because the hamiltonian vanishes when the
spins are aligned, the only energy is from the interface of this domain,
$ = J A$, with $A \sim R^{d-1}$ the area ($d$ is the number of spatial
dimensions).  This domain contributes to the partition function as
\beq
Z \simeq V \; e^{- J A/T} \;  .
\label{e3}
\eeq
The factor of volume in the measure, $V \sim R^d$,
is from the entropy of moving the domain
around.
Peierl's argument is that in more than one dimension, at low temperatures
the factor of $R^{d-1}$ in the exponential always wins over the $R^d$
in the measure: domains are exponentially suppressed, and the
symmetry is spontaneously broken.
In this view, it is not apparent how the symmetry is restored at
a finite temperature; this happens because when domains are common
they interact with one another, and this lowers their total energy [\ref{r3}].

\zn interfaces can be constructed in hot gauge theories.  In three
spatial dimensions the interface tension between two \zn domains is
[\ref{r4}]
\beq
J \simeq \frac{4 (N-1) \pi^2}{3 \sqrt{3N}} \,
\frac{T^3}{g} \; .
\label{e4}
\eeq
The detailed origin of the interface tension is a story in itself.
This result is computed semiclassically, and so is valid at high temperature,
where the running coupling is small.  The interface tension
is the stationary point of an effective action, and represents a
balance between classical and (certain) one loop effects.  Because of this
balance, $J$ is proportional not to $1/g^2$,
as is standard for instanton type solutions, but to $1/g$.

For my purposes here, however,
the powers of $g$ are inconsequential --- what matters is that
the mass dimension is set entirely by powers of temperature, and not
(say) by the renormalization mass scale.
It is then obvious why the symmetry is broken at high instead of
low temperature: the effective spin spin coupling is not constant
(as for the Ising model), but increases with temperature,
growing as $\sim T^3$ at large $T$.
The contribution of a domain to the partition function is
$Z \sim exp ( - \# T^2 R^2)$ at large $R$ and $T$: interfaces are rare
at high temperature, and become common only as the temperature
decreases.

What happens at large $N$, when $g^2 N$ is held fixed?  The usual free energy
of a gas of gluons is proportional to $N^2$, so it is natural to expect the
same for the spin spin coupling $J$.  Instead, one sees from
(\ref{e4}) that if $g^2 N$ is held fixed at large $N$, that $J \sim N$
and not $N^2$.  But remember that the \zn vacua are for phases
$\phi_j = 2 \pi j/N$, so if two \zn domains differ by a finite amount in
$j$, the change in phase is only of order $1/N$.  For two \zn domains
to differ by a finite phase, the difference in $j$ must be of order $N$.
The total energy for such an interface is of order $N J$, which is of
order $N^2$, as expected.

There is an amusing corollary to this picture.   Consider the above argument
at low temperatures.  If the interface tension is nonzero as $T \rightarrow
0$, then Peierl's argument implies symmetry breaking about zero
temperature.
Assuming that the symmetry is unbroken about zero temperature,
then for the picture of \zn domains to apply it is necessary for
the interface tension to {\it vanish} at least linearly with temperature,
$J \sim T$ as $T \rightarrow 0$.

I end this section with a matter of definition.
I propose that the vacuum expectation value of the Wilson line is
related to the free energy of a test quark, $F_q$, as
\beq
\langle L(x) \rangle \; = \; e^{- F_q/T + i \phi_j} \;  .
\label{e5}
\eeq
The phase factor $\phi_j$ depends upon which \zn domain the theory
falls into.  This relation is normally written without the phase factor
[\ref{r2}].
But then it is only consistent in the trivial phase, $\phi_j = 0$;
otherwise, (\ref{e5}) implies that when $j \neq 0$, the free energy is complex,
which is absurd.

The phase factor in (\ref{e5}) can be understood from the definition of
the underlying partition function.  To summarize: the Polyakov line
represents
not just the ``mass'' of the test particle (which at finite temperature is
the free energy), but more generally, the propagator,
$L \sim \bar{\psi}(\beta) \psi(0)$.  Now
the gauge field, which resides in the adjoint representation, is
invariant under the aperiodic gauge transformations mentioned previously.
Fields in the fundamental representation,however, do change:
while $\psi(0) \rightarrow \psi(0)$,
$\bar{\psi}(\beta) \rightarrow exp(i \phi_j) \bar{\psi}(\beta)$.  This
gives exactly the transformation expected for the Wilson line.
More generally, one can show that the transformations of the
Polyakov line under \zn transformations reflect the invariance of the
states of the (purely gluonic) theory under global
\zn gauge rotations [\ref{r5}].

\vspace{.25in}
\noindent{\bf 2. Singularity structure of gauge variant propagators}
\vspace{.125in}

In this section I conjecture what the possible singularity structure
of gauge variant propagators are like in covariant gauges at nonzero
temperature.  The motivation is recent work by Baier, Kunstatter, and
Schiff [\ref{r6}],
who showed that special care must be taken in covariant gauges.

I first give a simple argument for the singularity structure of gauge
variant propagators at zero temperature.
Consider $QED$ in covariant gauge, with gauge fixing parameter $\xi$.
Ignoring Dirac structure, at
one loop order the fermion self energy for a fermion of mass $m$
has the form
\beq
\Sigma^{1 \, loop}(\omega,k) \sim
\; c \, e^2 \; (\omega - E_k) \;
\int^m_{\omega-E_k} \frac{dk}{k} \; = \;
(\omega - E_k) \; c \, e^2 \;
ln \left( \frac{m}{\omega - E_k} \right) \; .
\label{e6}
\eeq
The external energy is assumed to be
near the mass shell, $\omega \sim E_k$, and $c = (3-\xi)/16 \pi$.
The logarithmic divergence is usually cut off by introducing a photon
mass or some such.  Instead, let me leave it as is, and assume that the
effects of all higher order is simply to sum up the lowest order term
into an exponential.  Then the logarithmic divergence at one loop order
becomes a branch point:
\beq
\Delta^{-1}(\omega,k) = \omega - E_k +
\Sigma^{1 \, loop}(\omega,k) + \ldots
\; \sim \; (\omega - E_k)^{1 + c \, e^2} \; .
\label{e7}
\eeq
While the above is just a guess, the result is correct:
the correct singularity structure for the fermion propagator
in $QED$ is a branch point, where the strength of the singularity is gauge
dependent [\ref{r7}].

At nonzero temperature I simply assume that the form is similar to that
at zero temperature, with the addition of the appropriate factors for
a thermal distribution:
\beq
\Sigma^{1\, loop}(\omega,k) \sim
(\omega - E_k) \; \xi \, \tilde{c}\, e^2
\; \int^T_{\omega-E_k} \frac{dk}{k}
(1 + 2 n(k) ) \; .
\label{e8}
\eeq
Here $n(k) = 1/(exp(k/T)-1)$ is the Bose-Einstein
statistical distribution function, and
$\tilde{c}$ is a computable, nonzero constant.
Note that at nonzero temperature
the gauge dependent term in (\ref{e8})
is {\it strictly} proportional to
$\xi$, and vanishes if $\xi$ does.  This is because at nonzero temperature,
the physical longitudinal and transverse modes acquire thermal ``masses''
which themselves cut off the infrared divergences near the mass shell.
The sole exception to this are gauge dependent terms in covariant gauges;
there the poles, massless at tree level, remain massless
to any order in perturbation
theory.  Also, I assume the upper limit on any infrared divergence is now
set by the temperature.
It is then easy to see that for nonzero temperature, since the Bose--Einstein
distribution function behaves as $n(k) \sim T/k$ at small k, that
the integral for the self energy now has a linear divergence,
\beq
\Sigma^{1\, loop}(\omega,k)
\sim \; (\omega - E_k) \; \xi \, \tilde{c} \, e^2
\; \left( \frac{2 \, T}{\omega - E_k} \right)\; .
\label{e9}
\eeq
This integral is precisely that studied by Baier, Kunstatter, and Schiff
[\ref{r6}]:
the factors of $\omega - E_k$ appear to cancel, giving a gauge dependent
contribution even on the mass shell.
Baier, Kunstatter, and Schiff showed that this term
has an imaginary as well as a real part, and so appears
to give a gauge dependent contribution to the damping rate --- which Braaten
and I claimed was gauge invariant [\ref{r8}].

Rebhan was the first to point out that
the above integral is infrared singular [\ref{r9}];
if computed with an infrared regulator, the term vanishes on the mass shell.
This is simply because at one loop order
there is a delicate balance between
$\omega - E_k$ upstairs against one over the same factor from
the integral.  Any infrared regulator will tend to smooth out the
integral, so the $\omega - E_k$ upstairs dominates, and $\Sigma^{1 \, loop}$
vanishes on the mass shell, $\omega = E_k$.

My purpose here is merely to note that this is {\it very} special to one
loop order and fails at higher order.  Indeed,
let us assume --- as is true at zero temperature --- that I can sum up
the most infrared singular terms by exponentiation of the one loop result,
to give
\beq
\Delta^{-1}(\omega,k) \sim
(\omega - E_k) \; exp\left( 2 \xi
\tilde{c} e^2 T \; \frac{1}{\omega - E_k} \right)
\; .
\label{e10}
\eeq
Because of the power like infrared singularities at nonzero temperature,
instead of a branch point singularity there appears to be an {\it essential}
singularity.

I do not have a grand moral from this calculation.  Probably for reasons
of prejudice as much as anything else,
I believe that the position of the
singularity in a gauge variant propagator is a gauge invariant quantity.
Certainly the general proof of Kobes, Kunstatter, and Rebhan
[\ref{r10}] does not
depend on the detailed nature of the singularity
in the inverse propagator: only that it vanishes on the mass shell.
But the tricks that work at one loop order probably will not be good
enough beyond that.  A more general methodology is needed.

\vspace{.25in}
\noindent{\bf 3. The problem of viscosity}
\vspace{.125in}

It's really quite surprising that it's so difficult to compute damping
rates in hot gauge theories.
Inevitably, this has bred a feeling
of ennui and despair:
after all, what are the damping rates good for?  How do they
affect measurable quantities?

One class of quantities which depend crucially on the damping rates are
the transport coefficients [\ref{r11},\ref{r12}].
In this section I repeat calculations of Ilyin et al. [\ref{r11}]
in order to emphasize an apparent contradiction.
Consider the Kubo
formula for
the shear viscosity, $\eta$; the bulk viscosity,
$\zeta$, can be computed from $\eta$.
The Kubo formula for $\eta$ is [\ref{r11}]
\beq
\eta \;=\; \frac{1}{5} \; \lim_{E \rightarrow 0} \;
\frac{1}{E} \; {\rm Im} \; D(-iE+0^+,0)	\; .
\label{e11}
\eeq
$D(p^0,\vec{p})$ is the Fourier transform
of a two point function of the stress
energy tensor $T^{i j}$:
$D(\tau, \vec{x}) \;=\; < \pi^{ij}(\tau,\vec{x}) \; \pi^{ij}(0,0) >$,
where $\pi^{ij} = T^{ij} -  T^{kk} \; \delta^{ij}/3$.
The discontinuity is computed after analytic continuation from
euclidean $p^0$ to $p^0 = - i E + 0^+$.

At lowest order the two point function of $T^{ij}$ is given
by the exchange of two transverse gluons,
\beq
D(p^0,0) \;=\; \frac{16}{3}
\; Tr \; \Delta_{t}(k^0,k) \Delta_{t}(p^0 -k^0,k)
\bigg(7 k^4 \;-\;  10 k^2 k^0 (p^0-k^0)
\;+\; 7 (k^0)^2 (p^0 -k^0)^2 \bigg) \; .
\label{e12}
\eeq
Here $k = |\vec{k}|$,
$Tr = T \sum_{k^0}  \int d^3 k/(2 \pi)^3$
is the integral over the loop momentum,
and $\Delta_t$ is the propagator for transverse gluons.
The four powers of momenta arise because
each $T^{ij}$ brings in two derivatives
of the gauge field.  Since the stress energy tensor is gauge variant,
gauge variant modes do not contribute to (\ref{e12}).
The contribution of the plasmon modes is neglected.  This is
allowed because momenta of order $T$ dominate, and in this range
the plasmon modes
are negligible (their residue is exponentially small).

Using the spectal representation of the transverse propagator,
\beq
\Delta_t(k^0,k) = \int^{1/T}_{0} d\tau \; e^{ik^0 \tau}
\int^{+\infty}_{-\infty} d\omega \; \rho_t(\omega,k) \left(
1 + n(\omega) \right) e^{- \omega \tau} \; ,
\label{e13}
\eeq
where $\rho_t(\omega,k)$ is the spectral density for transverse gluons,
and $n(\omega)$ the Bose--Einstein statistical distribution function,
\beq
\eta^{1 \, loop}_{gluon} \;=\; \frac{8}{15 \pi T}
\int_0^{\infty} k^2 dk \int^{+\infty}_{-\infty}  d\omega \;
n(\omega) (1 + n(\omega)) \rho_t^2(\omega,k) \;
\bigg(7 k^4 \;-\; 10 k^2 \omega^2 \;+\; 7 \omega^4 \bigg)
\label{e14}
\eeq
The spectral density is assumed to be a Breit--Wigner form,
\beq
\rho_t(\omega,k) \; = \; \frac{1}{2 \pi k}
\bigg( \frac{\gamma_{t}(k)}{(\omega - k)^2 + \gamma_{t}(k)^2}
\; + \; \frac{\gamma_{t}(k)}{(\omega + k)^2 + \gamma_{t}(k)^2}
\bigg) \; ,
\label{e15}
\eeq
where $\gamma_{t}(k)$ is the damping rate for a transverse gluon
of momentum $k$.  Then
\beq
\eta^{1 \, loop}_{gluon} \; = \; \frac{8}{15 \pi^2 T}
\; \int^{\infty}_{0} dk \; n(k) ( 1 + n(k))
\; \frac{k^4}{\gamma_{t}(k)} \; = \;
\frac{32 \pi^2}{225} \frac{T^4}{\gamma_{t}} \; .
\label{e16}
\eeq

The contribution of quarks can be computed similarly.
For three flavors of massless quarks, the total result in
$QCD$ is
\beq \eta^{1 \, loop} \; = \; \frac{\pi^2 \; T^4}{225} \; \left(
\frac{32}{\gamma_t} \; + \; \frac{63}{\gamma_+} \right) \; .
\label{e17} \eeq
In this expression $\gamma_t$ and $\gamma_+$ are the damping
rates for gluons and quarks with momenta of order $T$, which in that
range are independent of momenta.
Computation shows that these damping rates are of order $g^2 T$,
so in all
\beq
\eta^{1 \, loop} \; \sim \; \frac{T^3}{g^2} \; .
\label{e18}
\eeq
The exact value for $\eta^{1 \, loop}$ can be read off by using
(\ref{e17}) and the damping rates in ref. [\ref{r13}].  What I am
interested in here are the powers of $g$.

At first sight it might seem odd
that the shear viscosity diverges in the limit of a free
theory, $g \rightarrow 0$.  In fact this is a familiar feature
of the shear viscosity: it is infinite for an ideal gas because
if an ideal gas doesn't start out precisely in thermal equilibrium,
without interactions there is nothing to drive it there.

The problem of viscosity is the following.  The stress
viscosity can also be calculated using kinetic theory [\ref{r12}].
As is
standard in kinetic theory, the result for $\eta$ is one over
the cross section.  At lowest order the amplitude for the scattering
of two gluons going into two gluons is of order $g^2$, and so
\beq
\eta^{transport} \; \sim \; \frac{T^3}{g^4} \; .
\label{e19}
\eeq
Much work has gone into computing the subleading
logarithms in (\ref{e19});
Baym et al. [\ref{r12}] pointed out that the scale is set not by
magnetic mass (as first thought) but by Debye frequencies, of
order $g T$.  But forget the logarithms --- even the
{\it powers} of $g$ don't match up between (\ref{e18}) and
(\ref{e19}).

This discrepancy is special to hot gauge theories.
Consider a scalar field theory with four point coupling $\lambda$.
The damping rate first enters at two loop order,
$\gamma \sim \lambda^2$, so there one expects that both Kubo and transport
theory give $\eta \sim 1/\lambda^2$ [\ref{r14}].

It was pointed out to me first by Gordon Baym,
and later by Eric Braaten,
that $\eta^{1 \, loop}$ is incomplete.  This is also discussed
recently by Jeon [\ref{r14}].
The above calculation represents the
resummation of an infinite set of diagrams, which are self energy
corrections on two hard lines in a loop.  Besides this diagram,
there are also ladder diagrams, in which soft gluons
are emitted, all parallel to one another, from the top to the bottom
hard line in the loop.  One can easily show by power counting that
these diagrams are singular as $\gamma \rightarrow 0$,
and so contribute to the stress viscosity.

{\it If} the kinetic theory calculation is correction, then the
ladder diagrams must dominate, and give $\eta \sim 1/g^4$.  Then
$\eta^{1 \, loop}$ would be down by $g^2$.  Using the Kubo formula
would just be a dumb way of doing things.

{\it Otherwise}: the result is $\eta \sim 1/g^2$, with both ladder diagrams
and $\eta^{1 \, loop}$ contributing at the same order; then there is
more to the kinetic theory of
ultrarelativistic plasmas than first meets the eye.

I suspect that the former is most probably true.  But in either case,
one can {\it NOT} conclude that the damping rates are
irrelevant; they are certainly part
of the answer for the transport coefficients.

Lebedev and Smilga [\ref{r15}] have studied the contribution of the
damping rate to the photon self energy in hot $QED$.
They find that in the sum of the hard loop,
similar to $\eta^{1 \, loop}$,
and ladder diagrams, that the damping rate cancels.
While the damping rate cancels in the photon self energy, this does not
imply that it does so in all quantities.
Certainly there is a stress viscosity, computable (at least in
principle) from the Kubo formula.

I conclude this section by showing that once one has the stress viscosity
in hand, the bulk viscosity follows easily in a surprising way.
The standard relation between the bulk and stress
viscosities is
\beq
\zeta  \; = \; \left( 1  \; - \; 3 \;
\frac{\partial p}{\partial \epsilon} \right)^2 \; \frac{5}{3} \; \eta \; .
\label{e20}
\eeq
Here $p$ is the pressure and $\epsilon$ the energy
of the gas at a temperature $T$; calculation at next to leading order
shows
\beq
p \; = \; T^4 (a_1 - a_2 g^2 + \ldots) \; ,
\label{e21}
\eeq
where $a_1$ and $a_2$ are constants which depend upon the number of
colors and flavors [\ref{r16}].  The energy is given from the pressure by
$\epsilon = T^2 \partial(p/T)/\partial T$.  If the coupling constant
is considered as a fixed parameter, then it is easy to see that even
with the term $a_2$ in (\ref{e21}), that as for an ideal
gas $\partial p/\partial \epsilon = 1/3$, and the bulk viscosity
vanishes.

Of course the coupling is not a fixed parameter:
it ``runs'', and thereby develops a nontrivial dependence on
temperature.  With $c_1$ the lowest order coefficient of
the $\beta$--function, $\partial g/\partial ln(T) = - c_1 g^3$
($c_1 = (11 - 2N_f/3)/(16 \pi^2)$ for $N_f$ massless flavors in
$SU(3)$ color),
\beq
\zeta \; = \; \left(\frac{5 a_2 c_1}{6 a_1} g^4 \right) \; \frac{5}{3}
\; \eta \; = \; \frac{273,375}{1,478,656} \; \frac{g^8}{\pi^8} \eta
\; .
\label{e22}
\eeq
The number refers to three massless flavors in $SU(3)$ color.

The appearance of the lowest order coefficient of the $\beta$--function
isn't a complete surprise.
For an ideal gas, or any system in which the trace of the stress energy
tensor vanishes, $\partial p/\partial \epsilon = 1/3$.
As is well known, for massless theories the conformal anomaly implies
that the trace of the stress energy
tensor is proportional to the $\beta$--function, which is why
it enters into the relationship between the two viscosities.

For practical purposes the bulk viscosity is usually neglected relative
to the stress viscosity.  Certainly in the strict perturbative regime
$\zeta$ is negligible relative to $\eta$.  But the peculiar ratio
of integers in front of $\eta$ is about $1/5$; thus $\zeta$ is an
{\it extemely} sensitive function of $g$.  As the
coupling varies from $g \sim 1$ to $g \sim 3$, $\zeta$ varies from
essentially zero to a value commensurate with
$\eta$.  Perhaps this enormous
sensitivity has phenomenological consequences; after all,
$g \sim 3$ is still $\alpha = g^2/4 \pi$ of order one.

In summary, while I cannot point to a direct experimental probe
of the damping rates, they are crucial in sorting out the
transport coefficients, and thereby deserving of our attention.

I thank Randy Kobes and Gabor Kunstatter for a
most stimulating and enjoyable workshop;
especially those wonderful lunches on the
roof of the Museum of Art!
I also thank Ben Svetitsky, for correspondence which produced
the discussion in sec. I; Rolf Baier, for discussions at the
workshop which inspired sec. II; and lastly,
Eric Braaten, with whom the work in sec. III was done in
collaboration.  This work
was supported in part by the U.S. Department of Energy under
contract DE--AC02--76CH00016.

\vspace{.25in}
\noindent{\bf 4. References}
\newcounter{nom}
\begin{list}{[\arabic{nom}]}{\usecounter{nom}}
\item
A. M. Polyakov, \phl{72B}{477}{78};
L. Susskind, \prd{20}{2610}{79}.
\label{r1}
\item
L. D. McLerran and B. Svetitsky, \prd{24}{450}{81};
B. Svetitsky, Physics Reports {\bf 132}, 1 (1986).
\label{r2}
\item
For a modern discussion, see, e.g., S.-K. Ma, {\it Statistical Mechanics}
(World Scientific Publishing Co., Singapore, 1985).
\label{r3}
\item
T. Bhattacharya, A. Gocksch, C. Korthals-Altes, and R. D. Pisarski,
\prl{66}{998}{91}; \nup{B383}{497}{92}.
\label{r4}
\item
A. Gocksch and R. D. Pisarski, BNL preprint BNL-GP-1/93 (January, 1993).
\label{r5}
\item
R. Baier, G. Kunstatter, and D. Schiff, \prd{45}{4381}{92};
Bielefeld preprint BI-TP 92/19 (June, 1992).
\label{r6}
\item
Relativistic Quantum Theory, E. M. Lifshitz and L. P. Pitaevskii,
(Pergamon Press, London, 1977); sec. 116, especially (116.18).
\label{r7}
\item
E. Braaten and R.D. Pisarski,
\prl{64}{1338}{90}; \nup{B337}{569}{90}.
\label{r8}
\item
A. Rebhan, CERN preprint CERN-TH-6434-92; see, also:
E. Braaten and R. D. Pisarski,
\prd{46}{1829}{92};
H. Nakkagawa, A. Niegawa, and B. Pire,
Ecole Polytechnique preprints A156-0292 (February, 1992) and
A195-0992 (September, 1992), and ref. [\ref{r6}].
\label{r9}
\item
R. Kobes, G. Kunstatter, and A. Rebhan, \prl{64}{2992}{90};
\nup{B355}{1}{91}.
\label{r10}
\item
D. N. Zubarev, {\it Nonequilibrium Statistical Thermodynamics}
(Plenum Publishing Co., New York, 1974);
A. Hosoya, M.-A. Sakagami, and M. Takao, \anp{154}{229}{84};
R. Horsley and W. Schoenmaker, \nup{B280}{716}{87};\nup{B280}{735}{87};
\prl{57}{2894}{86};
F. Karsch and H. W. Wyld, \prd{35}{2518}{87};
S. V. Ilyin, A. D. Panferov, and Yu. M. Sinyukov,
\phl{B227}{455}{89};
S. V. Ilyin, A.D. Panferov, Yu.M. Sinyukov, S.A. Smolyansky,
and G.M. Zinovjev, Kiev preprint ITP-89-6E.
\label{r11}
\item
S.-P. Li and L. McLerran, \nup{B214}{417}{83};
A. Hosoya and K. Kajantie, \nup{B250}{666}{85};
U. Heinz; \anp{161}{48}{85}; \anp{168}{148}{86};
P. Danielewicz and M. Gyulassy, \prd{31}{53}{85};
S. Gavin, \nup{A435}{826}{85};
H. T. Elze, M. Gyulassy, D. Vasak, \nup{B276}{706}{86};
M. Mizutani, S.Muroya, and M. Namiki, \prd{37}{3033}{88};
H.--Th. Elze and U. Heinz, \phr{183}{81}{89};
G. Baym, H. Monien, C.J. Pethick, and D.G. Ravenhall, \prl{64}{1867}{90};
P. Bozek and M. Ploszajczak, \prd{41}{634}{90};
St. Mr\'{o}wczy\'{n}ski and P. Danielewicz, Regensburg preprint TPR-89-34;
M. H. Thoma, \phl{B269}{144}{91};
D. W. von Oertzen, \phl{B280}{103}{92}.
\label{r12}
\item
R. D. Pisarski, \prl{63}{1129}{89};
V. V. Lebedev and A. V. Smilga, \anp{202}{229}{90};
\phl{B253}{231}{91}; C. P. Burgess and A. L. Marini, \prd{45}{R17}{92};
A. Rebhan, \prd{46}{482}{92};
T. Altherr, E. Petitgirard, and T. del Rio Gaztelurrutia,
Annecy preprint ENSLAPP-A-378/92 (April, 1992);
R. D. Pisarski, Brookhaven preprint (Dec., 1992).
\label{r13}
\item
S. Jeon, Univ. of Washington preprint UW/PT-92-03 (October, 1992).
\label{r14}
\item
V. V. Lebedev and A. V. Smilga, Physica A {\bf 181}, 187 (1992).
\label{r15}
\item
J. I. Kapusta, \nup{B148}{461}{79}.
\label{r16}
\end{list}
\end{document}